# আইনস্টাইন-পোডোলস্কি-রোজেন প্যারাডক্স ও বেল উপপাদ্য

উজ্জ্বল সেন

হরিশ-চন্দ্র রিসার্চ ইনস্টিটিউট, হোমি ভাভা ন্যাশনল ইনস্টিটিউটের অন্তর্গত, ছতনাগ রোড, ঝুঁসি, এলাহাবাদ ২১১ ০১৯, ভারতবর্ষ (ইন্ডিয়া)

**Abstract:** We will discuss here the Bell theorem, which shows that "locality" and "reality" are together inconsistent with quantum theory.

## সংক্ষেপে

এখানে আমরা বেল উপপাদ্য সম্পর্কে জানব। এই উপপাদ্যে দেখানো হয় যে "স্থানীয়তা" ও "বাস্তবতা" সম্বলিত যেকোন ভৌত তত্ত্ব, কোয়ান্টাম তত্ত্বের সঙ্গে সামঞ্জস্যপূর্ণ নয়।

## গোড়ার কথা

প্রথমেই বলে রাখা ভাল যে এই লেখা পড়ার চেয়ে জন বেলের পেপারটা [1] পড়া ঢের বেশি কার্যকরী। তাছাড়া মারমিনের সুন্দর সমীক্ষাটা [2] রয়েছে। হালফিলের এ বিষয়ে গবেষণা জানতে চাইলে ব্রুনেয়া, কাড়ালকান্তি, পিরোনিও, স্কারানি, ও ড়ীনা-র সমীক্ষাটা [3] দেখা যায়। জুকড়স্কির সাম্প্রতিক নিবন্ধটার [4] কথাও এখানে বলে রাখা যাক।

১৯৩৫ সালে, অ্যালবার্ট আইনস্টাইন, বরিস পোডোলস্কি, ও নাথান রোজেন, একটা রি-সার্চ পেপারে [5] দাবি করলেন যে কোয়ান্টাম মেকানিক্স "অসম্পূর্ণ"। ওনারা ধরে নিলেন যে "স্থানীয়তা" এবং "বাস্তবতা" নামের দুটো প্রেমিস সমস্ত ভৌত তত্ত্বকে মেনে চলতে হবে, এবং তাত্ত্বিকভাবে দেখালেন যে কোয়ান্টাম তত্ত্ব এই গোত্রের ভৌত তত্ত্ব নয়। ১৯৬৪তে জন স্টুআর্ট বেল দেখালেন যে স্থানীয়তা এবং বাস্তবতা যুগপৎ মেনে চলা সমস্ত ভৌত তত্ত্ব, কোয়ান্টাম মেকানিক্সের পরিসংখ্যানগত গণনাকে লঙ্ঘন করে।

আইনস্টাইন, পোডোলস্কি, ও রোজেন কন্টিনিউয়াস ভেরিয়েবল ব্যবহার করেছিলেন তাঁ-দের কল্পনার এক্সপেরিমেন্টে, যেখানে অন্যান্য জিনিসের সঙ্গে ডিরাক ডেল্টা "ফাঙ্কশনের" প্রয়োগ ছিল। বোম ও আহারোনভ [6] এটাকে কোয়ান্টাম স্পিনের ভাষায় ব্যক্ত করলেন। বেল, তাঁর পেপারে, এই ফাইনাইট-ডায়মেনশনল স্পেসের ভাষাতেই তাঁর সিদ্ধান্তগুলো দেন।

# কী ধরে নিচ্ছি

প্রথমেই, বাস্তবতা আর স্থানীয়তা নামের প্রেমিসগুলো বলতে আমরা ঠিক কী বলতে চাইছি, সেটা বুঝে নেওয়া যাক।

**বাস্তবতা বা "লুকনো রাশি" সংক্রান্ত প্রেমিস বলতে কী বুঝব:** এই প্রেমিস অনুযায়ী আমরা ধরে নেব যে কোনো বস্তুর ওপর তার কোনো বৈশিষ্ট্যের পরিমাপ করাতে নিতান্তই সেই বৈশিষ্ট্যের মান নির্ণয় করা হয়। সেই বৈশিষ্ট্যের সেই মান ওই বস্তুর ভেতর, পরিমাপ করার আগে থেকেই, বর্তমান ছিল - পরিমাপ করায় সেটা জানতে পেরেছি মাত্র। ওই বস্তুর স্টেটে (বিবরণে) কিছু লুকনো রাশি আছে বলে আমরা ধরে নেব, যে ভেরিয়েবলগুলো ওই বস্তুর সমস্ত পরিমাপযোগ্য বৈশিষ্ট্যের মান জানে। এই লুকনো ভেরিয়েবলগুলো প্রত্যেক এক্সপেরিমেন্টাল রানে আলাদা হতে পারে। আর আজ তারা লুকনো হলেও, ভবিষ্যতে প্রকাশ হয়ে যেতে পারে [7]।

**স্থানীয়তা বলতে আমরা কী বুঝব:** কোনো বস্তুর ওপর কোনো বৈশিষ্ট্যের পরিমাপ করাতে যে মান আমরা পাব, সেটা অন্য কোনো দূরে অবস্থিত বস্তুর ওপর কী বৈশিষ্ট্য পরিমাপ করা হচ্ছে, তার ওপর নির্ভর করবে না। তাহলে এখানে আমরা দেখতে পাচ্ছি যে স্থানীয়তা নামক প্রেমিসটা প্রযোজ্য হতে হলে, পুরো বস্তুটার অন্তত দুটো অংশ থাকতেই হচ্ছে। আর তাই, পুরো বস্তুটার ডায়মেনশন অন্তত চার হতে হচ্ছে।

ব্যাপারটা আরেকটু পরিষ্কার করার জন্যে একটা উদাহরণ দেওয়া যাক। আগেই বলা হয়েছে যে আমাদের এমন একটা বস্তু নিয়ে কাজ করতে হবে, যেটার অন্তত দুটো অংশ আছে। ধরা যাক এই দুটো অংশ, দুটো আলাদা গবেষণাগারে রয়েছে। এই দুই জায়গায় একজন করে গবেষক আছে, এবং তাদের নাম ধরা যাক আকবর আর বীরবল। এরা দুজনেই আমার থেকে বয়েসে বেশ ছোট, আর তাই আমি ওদের "তুমি" করেই সম্বোধন করছি। তাছাড়া সংক্ষিপ্ত হবে বলে, আমরা মাঝে-মাঝেই এদের $A$ আর $B$ বলেও ডাকব।

$A$ ও $B$-এর কাছে একটা বস্তুর অনেককটা প্রতিরূপ আছে - যাদের প্রত্যেকের দুটো করে অংশ আছে। কোনো একটা এক্সপেরিমেন্টাল রানে, ওরা এই সমষ্টি থেকে কোনো একটা প্রতিরূপ ব্যবহার করে।

বাস্তবতা বলতে আমরা বুঝব, যে প্রত্যেক প্রতিরূপের ওপর $A$ ও $B$ যা যা বৈশিষ্ট্য পরিমাপ করতে পারে, তাদের সবার মান, ওই প্রতিরূপের লুকনো ভেরিয়েবলের জানা। আকবর ও বীরবল পরিমাপ করায় তারা পরিমাপ করা বৈশিষ্ট্যের মান জানতে পেরেছে। কিন্তু সেই মান ওই বস্তুর ভেতর আগেই থেকেই ছিল। ঠিক যেমন আমি যদি এক গোছা ফুলের ভেতর থেকে না দেখে একটা ফুল টেনে নিই, এবং তারপর সেটা দেখি এবং জানতে পারি যে সেটা লালরঙা, তাহলে আমি দেখার আগেও ফুলের রংটা লালই ছিল - আমি সেটার দিকে চোখ ফেলায় সেই রংটা জানতে পেরেছি মাত্র।

স্থানীয়তা বলতে আমরা বুঝব যে আকবর তার ল্যাবরেটরিতে $A$ ও $B$-এর যৌথ মালিকানায় থাকা বস্তুটার কোনো একটা প্রতিরূপের $A$-র দিকের অংশটার ওপর কী বৈশিষ্ট্য পরিমাপ করবে, তা বীরবলের গবেষণাগারে সেই একই প্রতিরূপের অন্য অংশটার ওপর কোনো পরিমাপ করা বৈশিষ্ট্যের মান নির্ণয় করবে না। আরও নির্দিষ্ট ভাবে বলার জন্যে ধরা যাক যে আকবর তার ল্যাবে শুধুমাত্র $A_1$ অথবা $A_2$ বৈশিষ্ট্যগুলোর পরিমাপ করতে পারে। একইভাবে, বীরবল তার গবেষণাগারে শুধুমাত্র $B_1$ অথবা $B_2$ বৈশিষ্ট্যগুলোর পরিমাপ করতে পারে। স্থানীয়তা বলতে আমরা বুঝব যে আকবর তার ল্যাবে $A_1$-এর পরিমাপ করল না $A_2$-এর, তার ওপর বীরবলের পরিমাপ করা বৈশিষ্ট্যের মানের পরিবর্তন হবে না। অর্থাৎ কিনা আকবর ও বীরবলের কাছে যদি একটা করে ফুল থাকে, তাহলে আকবরের কাছে থাকা ফুলটা সে ছুঁয়ে দেখল না শুঁকে দেখল, তার ওপর বীরবলের তার ফুলটা শুঁকে দেখায় ভালো লাগা না-লাগা নির্ভর করবে না। আমাদের মনে রাখা ভাল যে এই ফুলের উদাহরণ ব্যবহার করে স্থানীয়তার ব্যাখ্যা দেওয়াতে একটু অতি-সরলীকরণ হয়ে যায় - কারণ, ফুলের কোমলতা এবং

গন্ধ, দুটো কমুটিং বৈশিষ্ট্য।

এখানে বলে রাখা দরকার যে বাস্তবতা বা স্থানীয়তা সম্পর্কে কোয়ান্টাম তত্ত্ব কোনো সাড়াশব্দ করে না। কোনো বস্তুর কোনো বৈশিষ্ট্য, যা কিনা এখনো পরিমাপ করা হয়ে নি, তার মান কী, সে সম্পর্কে কোয়ান্টাম মেকানিক্স সম্পূর্ণ নীরব [৪]। একইভাবে, আকবর কোনো এক এক্সপেরিমেন্টাল রানে $A_1$ মেপেছে এবং বীরবল $B_1$, কিন্তু সেই রানে যদি আকবর $A_2$ মাপত, তাহলে বীরবলের $B_1$ বৈশিষ্ট্য পরিমাপের ফল কী হতো, সে ব্যাপারেও কোয়ান্টাম তত্ত্ব একেবারে স্পিকটি-নট।

আমরা আগেই ধরে নিয়েছি আকবর $A_1$ আর $A_2$, এই দুই বৈশিষ্ট্যের কোনো একটার পরিমাপ করবে, আর বীরবল $B_1$ আর $B_2$, এই দুটো বৈশিষ্ট্যের কোনো একটা মাপবে। এবার আমরা এও ধরে নেব যে এই বৈশিষ্ট্যগুলোর প্রত্যেকটা শুধুমাত্র দুটো করে ফল দিতে পারে। বুলিয়ান রাশির মতো। যেমন কোনো একটা ফুলের রং হয় লাল, নাহয় লাল নয়। বা কাল সকালে হয় বৃষ্টি হবে, নতুবা হবে না। আর গাণিতিক সুবিধার্থে আমরা এই ফলদুটোকে $+1$ আর $-1$ ধরে নেব। প্রসঙ্গত, এই রকম বৈশিষ্ট্য কোয়ান্টাম মেকানিক্সে বর্তমান। যথা, একটা কোয়ান্টাম স্পিন-$1/2$ সিস্টেমের কোনো এক দিকের স্পিন পরিমাপ করতে গেলে আমরা যে অপারেটরের ব্যবহার করি, তাদের মান $+1$ আর $-1$ হয়, $\hbar/2$ ফ্যাক্টরটাকে যদি অগ্রাহ্য করা হয়।

## যা ধরে নিচ্ছি, তা থেকে কী পাচ্ছি

বাস্তবতা সম্পর্কের প্রেমিসটাকে ধরে নিয়ে আমরা বলতে পারি যে আকবর ও বীরবলের কাছে থাকা দু-অংশের বস্তুটার ওপর কোনো একটা এক্সপেরিমেন্টাল রানে আকবর এবং বীরবল যদি যথাক্রমে $A_1$ আর $B_1$ মাপে, তাহলে তাদের মান হবে $A_1(\lambda)$ আর $B_1(\lambda)$, যেগুলো কিনা পরিমাপগুলোর আগে থেকেই বস্তুটার লুকোনো রাশি $\lambda$-র ভেতর ছিল। এখানে, $\lambda$ চিহ্নিত করছে ওই এক্সপেরিমেন্টাল রানে যে সমস্ত বৈশিষ্ট্য আমরা পরিমাপ করতে পারতুম - কিন্তু হয়তো করিনি বা হয়তো করেছি - তাদের সবার মান নির্ণয় করার জন্যে যে সমস্ত লুকোনো রাশির প্রয়োজন, তাদের সমষ্টি। ওই একই এক্সপেরিমেন্টাল রানে, যদি বীরবল $B_1$ না মেপে $B_2$ মাপত, তাহলে ও সেই পরিমাপের ফল হিসেবে পেত $B_2(\lambda)$। নজর করার বিষয় যে আমরা এখানে ইতিমধ্যে স্থানীয়তা সংক্রান্ত প্রেমিসটাকেও ধরে নিলুম, কারণ আকবরের $A_1$-এর পরিমাপোত্তর বা পরিমাপপূর্ব মান, বীরবল $B_1$ মেপেছে না $B_2$, তার ওপর নির্ভর করেনি।

অন্য যে সমস্ত মাপজোকের কম্বিনেশন করতে পারে আকবর আর বীরবল, সেগুলোর নোটেশনও একইরকম হবে। সেগুলো আর আলাদা করে বলা হলো না এখানে।

যে এক্সপেরিমেন্টটা করা হচ্ছে, মানে যেটা আকবর-বীরবল যৌথভাবে করছে, সেটার জন্যে, এবং যে দু-অংশের বস্তুটা তাদের কাছে রয়েছে, তার জন্যে, লুকোনো রাশি $\lambda$-টা ধরা যাক একটা প্রব্যাবিলিটি ডিস্ট্রিবিউশন $\rho(\lambda)$ অনুযায়ী বিন্যস্ত।

বেল উপপাদ্য প্রমাণ করার জন্যে আমাদের প্রথমে বেল অসমীকরণ কী, সেটা জানতে হবে। এই অসমীকরণ লেখার জন্যে চারখানা কোরিলেটর ব্যবহার করতে হবে। এই কোরিলেটরগুলো আকবর আর বীরবলের বিভিন্ন পরিমাপগুলোর ভেতরের পারস্পরিক সম্পর্কের পরিমাণ নির্ণয় করবে। উদাহরণ স্বরূপ, ধরা যাক আকবর $A_1$ মেপেছে আর বীরবল $B_1$। এই জুড়ির কোরিলেটর হল $\langle A_1 B_1 \rangle$, যা কিনা এইভাবে লিখতে পারি:

$$\langle A_1 B_1 \rangle = \int d\lambda \rho(\lambda) A_1(\lambda) B_1(\lambda). \tag{1}$$

এই একইভাবে আমরা অন্য কোরিলেটরগুলোকে ব্যক্ত করতে পারি। এগুলো ব্যবহার করে

আমরা লিখতে পারি,

$$\langle A_1 B_1 \rangle + \langle A_1 B_2 \rangle + \langle A_2 B_1 \rangle - \langle A_2 B_2 \rangle$$
$$= \int d\lambda \rho(\lambda) \left[ A_1(\lambda) \left( B_1(\lambda) + B_2(\lambda) \right) + A_2(\lambda) \left( B_1(\lambda) - B_2(\lambda) \right) \right]. \quad (2)$$

ওপরের ইক্যোশনের বাঁদিকের এক্সপ্রেশনটাকে বেল অপারেটর বলা হয়েছে অনেক পেপারে। এখানে যে আমরা বাস্তবতা সংক্রান্ত প্রেমিসটাকে ব্যবহার করেছি, তা আর বলার অপেক্ষা রাখে না। এবং আমরা স্থানীয়তাকেও ব্যবহার করেছি, কারণ আকবর $A_1$ পরিমাপ করায় যে মান পেয়েছে, সেটা বীরবল $B_1$ মেপেছে না $B_2$, তার ওপর নির্ভর করেনি।

সমীকরণ (2)-এ, $B_1(\lambda) + B_2(\lambda)$ আর $B_1(\lambda) - B_2(\lambda)$, এই দুটো এক্সপ্রেশনের ভেতর একটা শূন্য হবে, আর অন্যটা $+2$ বা $-2$ হবে। এখানে আমরা ব্যবহার করেছি যে $B_1(\lambda)$ এবং $B_2(\lambda)$ শুধুমাত্র $+1$ বা $-1$ মান গ্রহণ করতে পারে।

তাহলে আমরা পেলুম,

$$\left| \langle A_1 B_1 \rangle + \langle A_1 B_2 \rangle + \langle A_2 B_1 \rangle - \langle A_2 B_2 \rangle \right| = 2 \left| \int d\lambda \rho(\lambda) A_j(\lambda) \right|, \quad (3)$$

যেখানে $j$ হয় 1 বা 2। কিন্তু $j$ যাই হোক না কেন, $|A_i(\lambda)| = 1$, আর তাই আমরা এই অসমীকরণে পৌঁছই:

$$\left| \langle A_1 B_1 \rangle + \langle A_1 B_2 \rangle + \langle A_2 B_1 \rangle - \langle A_2 B_2 \rangle \right| \leq 2. \quad (4)$$

এটারই নাম **বেল অসমীকরণ**। বেলের ১৯৬৪ সালের পেপারে ঠিক এই অসমীকরণটা ছিল না। এই আকারে এটাকে লেখেন ক্লাওজার, হর্ন, শিমনি, ও হোল্ট [9]। এই আকারের সমীকরণটাই বহুল প্রচলিত হয়েছে। এটাকে অনেক সময় CHSH বা বেল-CHSH অসমীকরণও বলা হয়। বেল অসমীকরণ প্রমাণ করে আমরা প্রায় চোদ্দ আনা কাজই সেরে ফেলেছি - বেল উপপাদ্য প্রমাণ করার কাজ।

এটা মনে রাখা একান্তই জরুরি যে বেল অসমীকরণ প্রমাণ করতে আমরা কোথাও কোয়ান্টাম তত্ত্বের ব্যবহার করিনি। এবং এই অসমীকরণ সেই সমস্ত ভৌত তত্ত্বের জন্যে সত্যি, যারা বাস্তবতা ও স্থানীয়তা মেনে চলে।

কিন্তু আমরা দেখতে চেষ্টা করতেই পারি যে কোনো দু-অংশ বিশিষ্ট বস্তুর কোয়ান্টাম স্টেট ও তার ওপরে করা কোনো বৈশিষ্ট্যেদের পরিমাপ, বেল অসমীকরণ মেনে চলে কিনা। ধরা যাক, আমরা সেই বস্তু হিসেবে নিলুম দুটো কোয়ান্টাম স্পিন-$1/2$, এবং তাদের ওপর কিছু স্পিন পরিমাপ করলুম। এও ধরে নিলুম যে সেই বস্তুটার কোয়ান্টাম স্টেট হল সিঙ্গলেট স্টেট, অর্থাৎ

$$\left| \psi^- \right\rangle = \frac{1}{\sqrt{2}} \left( \left| \uparrow_z \right\rangle \left| \downarrow_z \right\rangle - \left| \downarrow_z \right\rangle \left| \uparrow_z \right\rangle \right). \quad (5)$$

এখানে, $\left| \uparrow_z \right\rangle$ ও $\left| \downarrow_z \right\rangle$ যথাক্রমে একটা কোয়ান্টাম স্পিন-$1/2$ সিস্টেমের $z$-ডিরেকশনের স্পিন-আপ ও স্পিন-ডাউন স্টেট। আমরা যদি $A_1$, $A_2$, $B_1$, $B_2$-র পরিমাপগুলোকে যথাক্রমে $\hat{a}_1$, $\hat{a}_2$, $\hat{b}_1$, $\hat{b}_2$ দিকের স্পিন কম্পোনেন্টের পরিমাপ হিসেবে চিহ্নিত করি, যেখানে $\hat{a}_1$, $\hat{a}_2$, $\hat{b}_1$, $\hat{b}_2$ হলো চারটে থ্রী-ডায়মেনশনল ইউনিট ভেক্টর, তাহলে $A_1 = \vec{\sigma} \cdot \hat{a}_1$, ইত্যাদি হবে, যেখানে $\vec{\sigma}$ হলো তিনটে পাওলি ম্যাট্রিক্সের একটা ভেক্টর। তাহলে, সিঙ্গলেট স্টেটের জন্যে আমরা পাবো,

$$\langle A_1 B_1 \rangle = -\hat{a}_1 \cdot \hat{b}_1, \quad (6)$$

ইত্যাদি, আর তাই, যদি বেল অসমীকরণ সিঙ্গলেট স্টেটের জন্যে সত্যি হয়, পরিমাপ করা বৈশিষ্ট্যগুলোর জন্যে, তাহলে আমরা পাবো,

$$\left| \hat{a}_1 \cdot \left( \hat{b}_1 + \hat{b}_2 \right) + \hat{a}_1 \cdot \left( \hat{b}_1 - \hat{b}_2 \right) \right| \leq 2. \tag{7}$$

উল্লেখ্য যে আমরা এখানে কোয়ান্টাম মেকানিক্সের বর্ন বিধি ব্যবহার করেছি। সহজেই দেখা যায় যে ওপরের এই অসমীকরণ মোটেই সমস্ত $\hat{a}_1$, $\hat{a}_2$, $\hat{b}_1$, $\hat{b}_2$-এর জন্যে সত্যি নয়। ইউনিট ভেক্টরগুলোকে পোলার কোঅর্ডিনেটে লিখে একটু টানাহাঁচড়া করলেই পাওয়া যাবে যে ওপরের অসমীকরণের বাঁদিকটা শুধু দুয়ের ওপরেই ওঠে না, এমনকি $2\sqrt{2}$ পর্যন্ত যায়। সাই-রেলসন [10] দেখিয়েছেন যে এই $2\sqrt{2}$-ই হলো বেল-CHSH অসমীকরণের সর্বোচ্চ লঙ্ঘন। অর্থাৎ, অন্য কোনো কোয়ান্টাম স্টেট নিয়ে এবং অন্য কোনো বৈশিষ্ট্যের পরিমাপ করে অসমীকরণ (4)-এর ওই বাঁদিকের এক্সপ্রেশনটাকে $2\sqrt{2}$-এর বেশি কোনো ভ্যালুতে নিয়ে যাওয়া যাবে না।

সুতরাং আমরা দেখতে পাচ্ছি যে এমন দু-অংশ-বিশিষ্ট বস্তুর অবস্থা আছে এবং তার ওপর এমন বৈশিষ্ট্যের পরিমাপ আছে, যার গোটা ডেসক্রিপশন যদি কোয়ান্টাম তত্ত্ব অনুযায়ী করা হয়, তাহলে সেই পরিমাপগুলোর কোরিলেটররা বেল অসমীকরণ লঙ্ঘন করে। অর্থাৎ, কোয়ান্টাম তত্ত্ব এমন কোনো ভৌত তত্ত্বের সঙ্গে সামঞ্জস্যপূর্ণ হতে পারবে না যেটা বাস্তবতা ও স্থানীয়তা দুটোই মেনে চলে। এটাই **বেল উপপাদ্যের** বিবৃতি।

ওপরে এক জায়গায়, কোনো দুটো বৈশিষ্ট্য কম্যুটিং হয়ে গেলে সেটা অতি-সরল হয়ে যাবে বলে মন্তব্য করেছিলুম। এখানে দেখছি যে যদি আকবরের $A_1$, $A_2$ বা বীরবলের $B_1$, $B_2$ কম্যুট করে, তাহলে আমরা আর বেল অসমীকরণ লঙ্ঘন করতে পারব না, সে যে স্টেটই নিই না কেন।

## সত্যিকারের যন্ত্রে একটু গোলযোগ থাকেই

আমরা ধরে নিতে পারি যে সত্যিকারের কোনো এক্সপেরিমেন্ট করতে গেলে, সিঙ্গলেট স্টেটটা ঠিক সিঙ্গলেট হিসেবে আকবর আর বীরবলের মাপযোগ করার যন্ত্রে পৌঁছবে না। ঠিক কী অবস্থায় পৌঁছবে, সেটা নির্ভর করবে সেট-আপটা ঠিক কোন পরিমণ্ডলে থাকবে। এই পরিমণ্ডলের ছাঁচ ঠিক কী এবং সেই পরিমণ্ডলের সাথে আকবর-বীরবলের হেফাজতে থাকা দু-অংশের বস্তুটার ঠিক কী যোগাযোগ আছে, তা জানা অনেকসময়েই দেবতার অসাধ্য হয়ে ওঠে। এইরকম গোলযোগপূর্ণ পরিস্থিতির ক্ষেত্রে একটা বহুল-ব্যবহৃত মডেল হলো "হ্যাইট নয়েজ", যেখানে ধরে নেওয়া হয় যে সব গোলমাল মিলেমিশে গিয়ে এমন অবস্থা দাঁড়াবে যে আকবর আর বীরবল যে স্টেটটা পাবে, সেটা এইরকম হবে:

$$\varrho_W^p = p \left| \psi^- \right\rangle \left\langle \psi^- \right| + \frac{1}{4}(1-p)\mathbb{I}_2 \otimes \mathbb{I}_2. \tag{8}$$

এই স্টেটটাকে ভেয়ানা স্টেট [11] বলে। এই স্টেটটাকে যদি পজিটিভ হতে হয়, তাহলে $p$-কে $-1/3$ আর $1$-এর ভেতর থাকতে হবে। স্টেটটাকে যদি আমরা সিঙ্গলেট আর হ্যাইট নয়েজের (অর্থাৎ $(1/4)\mathbb{I}_2 \otimes \mathbb{I}_2$ স্টেটটার) প্রোব্যাবিলিস্টিক মিশ্রণ হিসেবে ভাবতে চাই, তাহলে অবশ্য $p$-কে ঋণাত্মক হতে দেওয়া যাবে না। এই $p$-কে অনেকসময় সেট-আপটার ভিজিবিলিটি বলা হয়। এখানে $\mathbb{I}_2$ হলো একটা টু-ডাইমেনশনল কমপ্লেক্স হিলবার্ট স্পেসের ওপরের আইডেন্টিটি অপারেটর, অর্থাৎ একটা $2 \times 2$ আইডেন্টিটি ম্যাট্রিক্স।

এখন সিঙ্গলেট পেলে আকবর-বীরবলের যন্ত্রপাতি বেল অসমীকরণের $2\sqrt{2}$ পর্যন্ত লঙ্ঘন দেখাতে পারে। ভেয়ানা স্টেটটা পেলে অবশ্যই আর অতটা পারবে না। কতটা পারবে, সেটা

নির্ভর করবে $p$ কত, তার ওপর। লঙ্ঘন শুরু হয় $2$ পরেই। যেহেতু $2\sqrt{2}$ রয়েছে $2$ থেকে কিছুটা দূরে, এটা বোঝাই যাচ্ছে $p = 1$ থেকে শুরু করে কিছুদূর পর্যন্ত এই লঙ্ঘন চলবে। একটু হিসেবনিকেশ করলেই দেখা যাবে, যে এলাকায় লঙ্ঘন হবে, সেটা হলো

$$p \in \left(\frac{1}{\sqrt{2}}, 1\right]. \tag{9}$$

## গ্রীতে হেয়ামান

স্থানীয়তা প্রেমিস ছাড়াই, শুধুমাত্র বাস্তবতা আছে এমন সমস্ত ভৌত তত্ত্ব কোয়ান্টাম মেকানিক্সের সঙ্গে অসামঞ্জস্যপূর্ণ - এইরকম দাবি বেল উপপাদ্য প্রমাণের অনেক আগে থেকে ছিল। বেল তাঁর ১৯৬৪-র গবেষণাপত্রে এর উল্লেখ করেন। কিন্তু এই প্রমাণ সঠিক নয়। এই ভুল বোধহয় প্রথম চোখে পড়ে গ্রীতে হেয়ামানের [12, 13]। বেল নিজেও পরে এই ভুল লক্ষ্য করেন।

আসলে, বাস্তবতা এবং কোয়ান্টাম তত্ত্ব - এ দুটোর সঙ্গেই একসাথে সামঞ্জস্যপূর্ণ ভৌত তত্ত্ব, লিটারেচারে আছে। উদাহরণস্বরূপ, বোম মডেলের [14] কথা বলা যায়। অবশ্যই সেরকম সমস্ত তত্ত্ব নিশ্চিতরূপে স্থানীয়তা লঙ্ঘন করবে, বেলের উপপাদ্য অনুযায়ী।

টু-ডায়মেনশনল কোয়ান্টাম সিস্টেমের জন্যে একটা বাস্তবতা এবং কোয়ান্টাম তত্ত্বের সঙ্গে সামঞ্জস্যপূর্ণ ভৌত তত্ত্ব, বেলের ১৯৬৪-র পেপারে রয়েছে। অবশ্য সেখানে স্থানীয়তা আছে না নেই, সেই প্রশ্নেরই অর্থ নেই।

## পরিশেষের পরিবর্তে

পরিশেষের পরিবর্তে, প্রথমেই বলে রাখা যাক যে বেল অসমীকরণের বিভিন্ন সম্প্রসারণ হয়েছে। প্রত্যেকটা সম্প্রসারণ কোনো না কোনো নতুন দৃষ্টিকোণ এনেছে। কখনো হয়তো গবেষণাগারে বেল অসমীকরণের লঙ্ঘনে নতুন কোনো সহায়তা করেছে। কখনো হয়তো একটা নতুন ধারণা নিয়ে এসেছে। কখনো বা কোনো নতুন উপযোগ খুঁজে পাওয়া গেছে [3]।

দ্বিতীয়ত, গবেষণাগারে বেল অসমীকরণ লঙ্ঘন করাটা বেশ কষ্টসাধ্য। কোনো নতুন এক্সপেরিমেন্টাল নিজেরের ভেতর সাধারণভাবে কোনো না কোনো "লুপহোল" খুঁজে পাওয়া যায় [15]।

এবং সবশেষে বলে রাখা যাক যে বহু ধরণের রিসার্চ এরিয়া বেল উপপাদ্য দ্বারা প্রভাবিত হয়েছে। উদাহরণ হিসেবে এনট্যাঙ্গেলমেন্ট উইটনেসের বলা যায় - যা কিনা এই মুহূর্তে গবেষণাগারে এনট্যাঙ্গেলমেন্ট সনাক্ত করার অন্যতম ভাল উপায়। [এনট্যাঙ্গেলমেন্ট হলো এক ধরণের কোয়ান্টাম কোরিলেশন [16]।] ডিভাইস-ইন্ডিপেন্ডেন্ট ক্রিপ্টোগ্রাফি হলো আরেকটি জায়গা যেখানে বেল অসমীকরণের সুন্দর প্রয়োগ হয়েছে [17]।

## তথ্যসূত্রসমূহ